\documentclass[useAMS,usenatbib]{mn2e}
\usepackage[T1]{fontenc}
\usepackage{aecompl}
\usepackage{graphicx}
\usepackage{hyperref}

\newcommand{\mnras}{MNRAS}
\newcommand{\apj}{ApJ}
\newcommand{\apjl}{ApJL}
\newcommand{\aap}{A\&A}

\title{Satellite Infall and Mass Deposition on the Galactic Centre}
\author[Sof\'ia G. Gallego and Jorge Cuadra]{Sof\'ia G. Gallego$^{1,2}$ and Jorge Cuadra$^{1}$\thanks{E-mail:
gallegos@phys.ethz.ch (SGG); jcuadra@astro.puc.cl (JC)}\\
$^{1}$Instituto de Astrof\'isica, Pontificia Universidad Cat\'olica de Chile, 782-0436 Santiago, Chile
\\
$^{2}$Institute for Astronomy, Department of Physics, ETH Z\"urich, CH-8093 Z\"urich, Switzerland}
\begin{document}

\date{\today}

\pagerange{\pageref{firstpage}--\pageref{lastpage}} \pubyear{2016}

\maketitle

\label{firstpage}

\begin{abstract}
We model the infall of a $\sim2\times 10^5\mathrm{M}_{\odot}$ satellite galaxy on to the inner 200 parsec of our Galaxy, to test whether 
the satellite could perturb the gas previously on stable orbits in the central molecular zone (CMZ), as proposed by \citet{Lang13}.
This process would have driven a large gas inflow
around 10 Myr ago, necessary to explain the past high 
accretion rate onto the super-massive black hole, and the presence of young stars in the inner parsecs of the Galaxy. 
Our hydrodynamical simulations show a much smaller inflow of gas, not sufficient to produce the aforementioned effects.
 
\end{abstract}


\section{Introduction}

Sgr~A*, the super-massive black hole in our Galactic Center (GC) is characterized by its current very low luminosity \citep[see][for a review]{Genzel10}. However, there is evidence in the surrounding medium that suggests the GC was much more active in the past. The  \textquotedblleft Fermi Bubbles\textquotedblright \citep{Dobler10,Su10,Yang14, Ackermann14}, a pair of gamma-ray emitting bubbles located nearly perpendicular to the Galactic plane, and extending kiloparsecs away from the GC, are interpreted as the result of AGN-like and/or starburst activity around Sgr~A* \citep{Guo12, Yang12}.  
If AGN-like, it would imply that accretion on to the black hole was much stronger a few million years ago \citep{Zubovas12, Mou14}.

Recent mass supply is required to enable either  accretion or starburst activity. We observe evidence for this mass supply also in the form of young massive compact star clusters (the Arches, Quintuplet, and Central cluster), located within $30\, \mathrm{pc}$ from the GC \citep{Namekata09}. 
From the number of short-lived OB stars present
\citep{Figer03}, we can determine that an episode of star formation occurred within the central region of our Galaxy during the last several million years \citep{Mezger96} for the three clusters.

The most striking of these star clusters is the one at the very center of the Galaxy, located on top of Sgr~A*. 
Although it is currently accepted that the young stars formed in situ in one or more self-gravitating gaseous discs
\citep{Levin03, Nayakshin05, Paumard06, Nayakshin07},
it is not clear how those discs initially formed \citep[e.g.,][]{Cuadra08}.
Several numerical studies have explored the evolution of $\sim 10^5 \mathrm{M}_{\odot}$ giant molecular clouds that are captured by Sgr~A* and end up forming such discs
\citep{Bonnell08, Hobbs09, Alig13}. 
Recently \citet{Zubovas12}
proposed that the same process could enhance Sgr A*'s accretion and produce the
\textquotedblleft Fermi Bubbles\textquotedblright.

The most likely source for the gas is the Central Molecular Zone (CMZ), a region from $l = -1\deg$ to $l = 1.5\deg$, rich in molecular gas with a mean surface density of $200\,\rm{M_\odot pc^{-2}}$\citep{Morris97}. The main concentration of gas lies in a quasi ring-like structure of around $\sim180$ pc, as shown in the longitude---velocity ($lv$) diagrams
\citep{2006JPhCS..54...35R}.
The CMZ is produced by gas inflow from the rest of the Galaxy, through several angular momentum loss processes caused by a stellar bar potential which dominates the inner few kiloparsecs \citep{Zhao94}. 
The gas is initially in so-called X1 orbits, between the corotation radius and the Inner Lindblad Resonance (ILR) of the bar pattern.  As gas approaches the ILR, there is an innermost stable orbit inside of which the orbits become cusped or self-intersecting. Gas clouds experience compression and shocks near the cusps of these orbits, lose angular momentum, and plunge to the so-called X2 orbits inside. This is a family of closed and elongated orbits, with their long axes oriented perpendicular to the bar \citep{Morris96}.
Additional angular momentum is lost when infalling gas clouds collide with those already on X2 orbits, compressing and cooling the gas into molecular form. The accumulated molecular gas in the X2 orbits originates the observed CMZ ring.

The inflow of gas from the X1 orbits is mainly compensated by star formation and evaporation of clouds into the hot ($10^7 - 10^8\,$K) CMZ ring surroundings \citep{Koyama89,Yamauchi90,Sunyaev93,Koyama96}. The mean life of the molecular gas is around $(0.4 - 1) \times 10^9$ years, which is an order of magnitude shorter than the age of the Galaxy \citep{Morris97}. However, it is unclear the mechanisms which transport the gas further in towards the GC. Secondary effects, such as dissipation, gravitational instability, and magnetic viscosity, can then drive the gas further in, but at a slower rate \citep[e.g.][]{Fukuda00,Heller01,Morris96}.
 \citet{Lang13} (hereafter L13) recently proposed a novel mechanism, in which a satellite galaxy merger could have perturbed the CMZ orbits, driving gas into inner regions and compressing it beyond the density needed for massive star formation  \citep{Mihos96,Hopkins10}.
According to L13, the satellite must have had an initial mass of around $2\times10^8\rm{M_\odot}$,
falling from an orbit of initial eccentricity $e = 0.9$. By the time it reaches the CMZ the
remaining satellite mass would be about $2\times10^5\rm{M_\odot}$ in an almost coplanar orbit. From theoretical calculations, L13 estimates that the gravitational perturbation of the satellite remnant when reaching the CMZ region could produce a net infall of $10^6\rm{M_\odot}$ on a timescale of $\sim10$ Myr.

In this work, we test the effectiveness of this satellite-merger scenario by performing hydrodynamical simulations of the Milky Way CMZ perturbed by a compact satellite. In section 2 we describe the physical model and numerical setup used for our simulations. Section 3 presents the results obtained by exploring the parameter space.  Finally, in section 4 we draw our conclusions comparing with the initial L13 idea.

\section[]{Physical model \& numerical set-up}

We model the satellite infall scenario using the public Smoothed Particle Hydrodynamics (SPH) code Gadget-2 \citep{Springel05}.
We modified the code to include the bar-like gravitational potential of the inner region of the Galaxy, taken from \citet{Zhao94},
\begin{center}
\begin{equation}
 \Phi(r,\theta,\phi) = 4\pi G \rho_0 r_0^2 {\left( \frac{r}{r_0} \right) }^\alpha P(\theta,\phi),
\end{equation}
\end{center}
where $(r,\theta,\phi)$ are spherical coordinates fixed on the rotating bar, $P$ is the associated Legendre function,
$$P(\theta,\phi) = \frac{1}{\alpha(1+\alpha)}+\frac{Y(\theta,\phi)}{(2-\alpha)(3+\alpha)},$$ 
and $Y$ is a linear combination of spherical harmonic functions of the $l = 2$, $m =0,2$ modes,
$$ Y(\theta,\phi) = -b_{20} P_{20}(\mathrm{cos}\,\theta)+b_{22} P_{22}(\mathrm{cos}\,\theta) \mathrm{cos}\,2\phi.$$ 
   The parameter $b_{20}$ determines the degree of oblateness/prolateness while $b_{22}$ determines the degree of non-axisymmetry.

Based on previous work which shows the formation of a CMZ with this potential at around 200 pc \citep{Kim11} we chose the following parameters:
$\alpha = 1.75$, $b_{20}=0.3$, $b_{22}=0.1$, $\rho_0= 40 \,\mathrm{M}_{\odot}\, \mathrm{pc}^{-3}$, and $r_0=100\,\mathrm{pc}$. Assuming a solid body rotation at the scales of $<1$kpc \citep{Luna06,Kunder12},
the bar pattern speed is $63\, \mathrm{km}\, \mathrm{s}^{-1}\, \mathrm{kpc}^{-1}$. 
Given the above parameters, we obtain a bar with axis ratio of
$[1:0.74:0.65]$, an $X_1-X_2$ transition at $r\sim200\,\mathrm{pc}$, ILR at 660 pc, a corotation radius
of 2.8 kpc, and enclosed mass inside $200\, \mathrm{pc}$ of $10^9\, \mathrm{M}_{\odot}$.  

In addition to the gravitational force due to the potential above, centrifugal and coriolis forces were introduced in the code as we are working in the reference system of the rotating bar.

The gas of the CMZ is modeled as an ensemble of SPH particles.
Their initial conditions, namely periodic and closed orbits in the potential above, were obtained with a shooting technique. This starts with an assumption for the initial velocity in the x-y phase-space, in which the x-velocity component is zero on the semi-major axis of the orbit. The equations of motion are then integrated with a fourth order Runge-Kutta method on the reference system of the rotating bar. The initial velocity assumption is then optimized until obtaining a periodic orbit with a fractional precision of $10^{-7}$. 
The density distribution of the ring was produced following \citet{Ferriere07} space-averaged density profile, truncated between 130 and 230 pc,
\begin{equation}
 <n_{H}>_{\rm ring}(r,z)=  (106\, \mathrm{cm}^{-3})\, e^{ -\left(\frac{r-180}{L}\right)^4} \times 
 e^{-\left(\frac{|z|}{H}\right)^{1.4}},
\end{equation}
where $r$ and $z$ are the radial and vertical directions, respectively, and $L=55$ pc and $H=29$ pc are the corresponding lengthscales.
We run 2-d and 3-d simulations. The simulations were initially run without a satellite for 20 Myr in order to obtain a hydrodynamically-relaxed distribution of particles.  
The satellite was then introduced, modeled as a single `dark matter' particle with a softening of 10 pc, which mimics the effect of its expected non-zero size.

For our 3-d and 2-d fiducial simulations we used $500,000$ and $300,000$ particles respectively for the gas ring, with a total mass of $10^7\,\rm M_\odot$ and a 
fixed temperature of $1000\,\rm K$.
We do not include radiative cooling/heating nor any feedback processes, therefore the thermodynamics depends simply on the gas temperature.
Since the timescales we are modelling  are shorter than the viscous time, we do not model the gas transport through viscosity, but we do include the standard SPH treatment of shocks
through artificial viscosity, with a strength set to $\alpha = 0.8$, following the Monaghan-Balsara form (Monaghan and Gingold 1983; Balsara 1995) and using a 
viscosity limiter to prevent spurious angular momentum transport due to shear flows (Balsara 1995; Steinmetz 1996).

Considering these parameters and the rotation of the bar the Toomre stability parameter $Q$ is close to 1. The satellite has a mass of $2\times10^6\,\rm{M_\odot}$ (an order of magnitude higher than in L13 scenario), in a co-rotating and coplanar orbit, with its first impact to the CMZ around 1 Myr after the beginning of the simulation and initial radius and velocity of $300\, \rm{kpc}$ and $100\,\rm{km/h}$ respectively. The velocity is $20\%$ higher than the corresponding velocity for a periodic orbit at that distance, which permits the satellite to cross several times the CMZ.  Besides these fiducial simulations, we produced other models exploring the relevant parameter space, as detailed in next section.

\section{Results}

We explore the parameter space in terms of the gas mass, temperature, the satellite mass and initial orbit, and the bar pattern speed. Besides, we run control simulations without satellite for both 2-d and 3-d configurations. All simulations were run for 100 Myr, an order of magnitude longer than the satellite merger timescale proposed by L13.  
The models are listed in Table~1, showing also the amount of mass that reaches the inner 80 pc (a generous definition of the central region) after 20 and 40 Myr.  
Even after 40 Myr, the control simulations show little inflow, as expected for X2 orbits: in the two 2-d control simulations there was no measurable infall of gas after 40 Myr, for both resolutions we used, 300.000 (Figure 1 top left) and 1.1 million gas particles; in the 3-d control simulation $\sim10^3\,\rm M_\odot$ of gas reached the inner 80 pc of the GC after 40 Myr, which is explained by the natural vertical instabilities of the bar shaped potential. In the fiducial models (Figure 1, right panels), we see an increment of roughly twice the mass inflow with respect to the control runs.  This means that the satellite contribution is measurable, but still very small -- several orders of magnitude lower than the expected $10^6\,\rm M_\odot$ after 10 Myr from L13 estimations. 

Changing the satellite mass does not have a dramatic effect either. 
With a satellite mass of $2\times10^5\,\rm M_\odot$, the inflow is practically the same than in the control simulations; with a satellite mass of $2\times10^7\,\rm M_\odot$ the ring shape of the CMZ is noticeable perturbed, but the vast majority of the gas stays within its initial orbital radial range.  We did not test even higher satellite masses, as we would expect there would be more direct observational evidence for such an event on the tested timescale  \citep{Gilmore02}.
When the initial orbital parameters of the satellite are changed, such as counter-rotating or inclined orbits (Figure 2, right panel), the inflow of gas is similar 
to the fiducial at early stages and closer to the control simulations at 40 Myr and beyond. 
When changing the initial velocity to half its fiducial value we appreciate a slight increase of inflow, likely explained by a longer interaction between the satellite and the CMZ at each cross.

We analyzed the stability of the CMZ by modifying the temperature of the gas, which changes the Toomre parameter $Q$ to a value higher than 1 (higher temperature, more stable) 
or lower than 1 (lower temperature, less stable). We appreciate no significant changes when the CMZ is more stable.  
However, we notice a moderate increase of the inflow rate compared to the fiducial model when decreasing the gas temperature by 1/10. 

The overall small effect of the satellite perturbation is mostly explained by the stability of the CMZ ring sustained by the rotating bar, so we test the effect of changing its 
main property.   Observational studies set the pattern speed of our Milky Way bar/bulge between 30 to 60 $\mathrm{km}\, \mathrm{s}^{-1}\, \mathrm{kpc}^{-1}$ 
\citep{Gardner10, Minchev10, Wang12, Long13, Antoja14},
with the most recent Milky Way models and specially the BRAVA results leaning towards the low side \citep{Portail15, Li16}.  
Motivated by the latest results we test a final model with half the bar pattern speed ($30\, \mathrm{km}\, \mathrm{s}^{-1}\, \mathrm{kpc}^{-1}$). 
No significant increase in the infalling mass was found with respect to the previous simulations.

According to the literature, minor mergers have an important rol in the star formation history of galaxies, the evolution of galactic bulges, 
and on fueling active galactic nuclei \citep{Corbin00,Miralles14,Kaviraj14}, but little is known below kiloparsec scales. 
The results of our simulations are interesting in this context, and would imply that a single minor interaction
is not effective to drive inflow from central molecular zones when a bar is present, whereas this inflow may be
produced by bar instabilities themselves \citep{Fanali15,Dullo16,Lin16, Bonoli16}. A much larger exploration of the galaxy parameter space is needed to address this issue.

\begin{figure*}
\begin{center}
\includegraphics[width=3.3in]{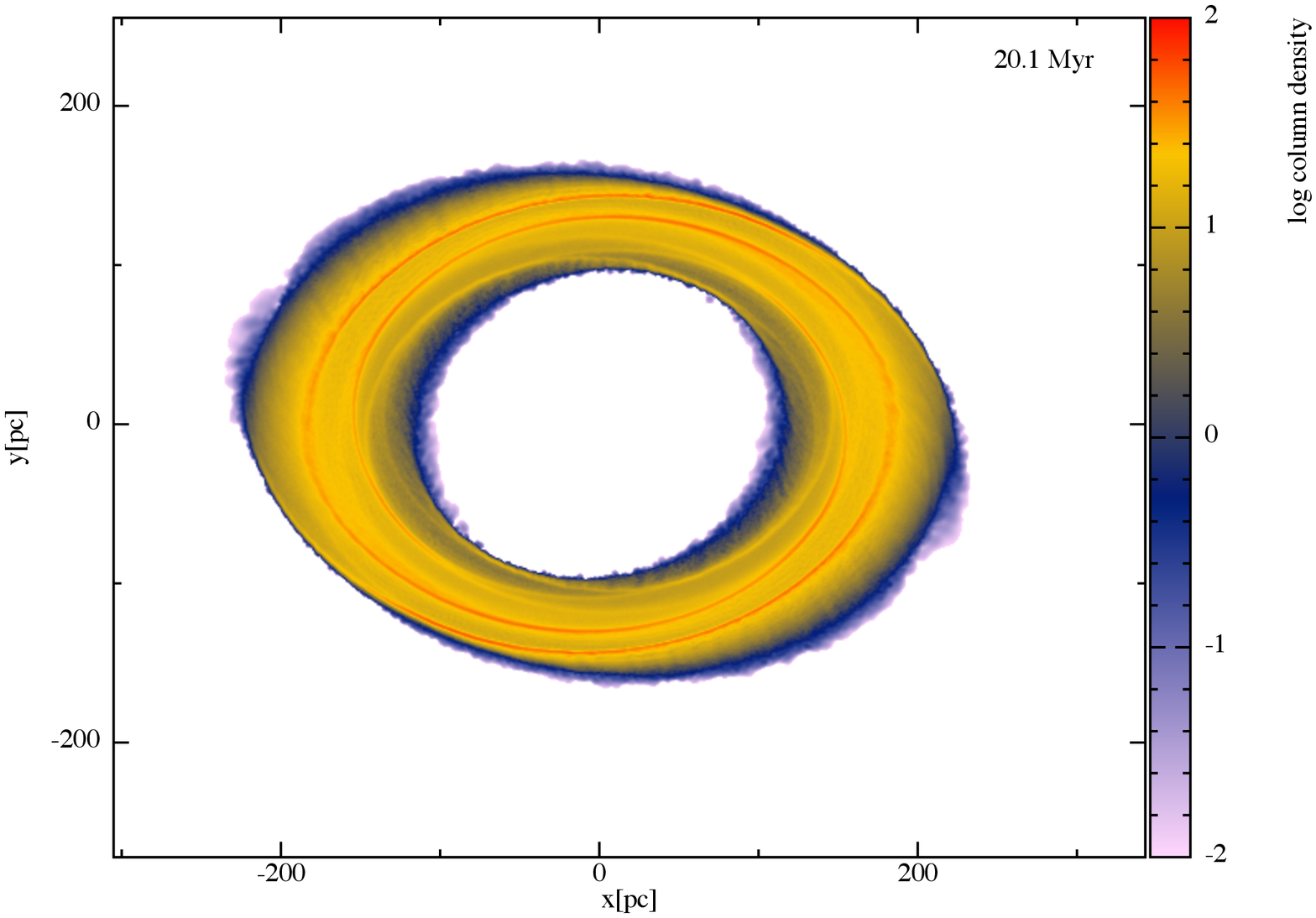} 
\includegraphics[width=3.3in]{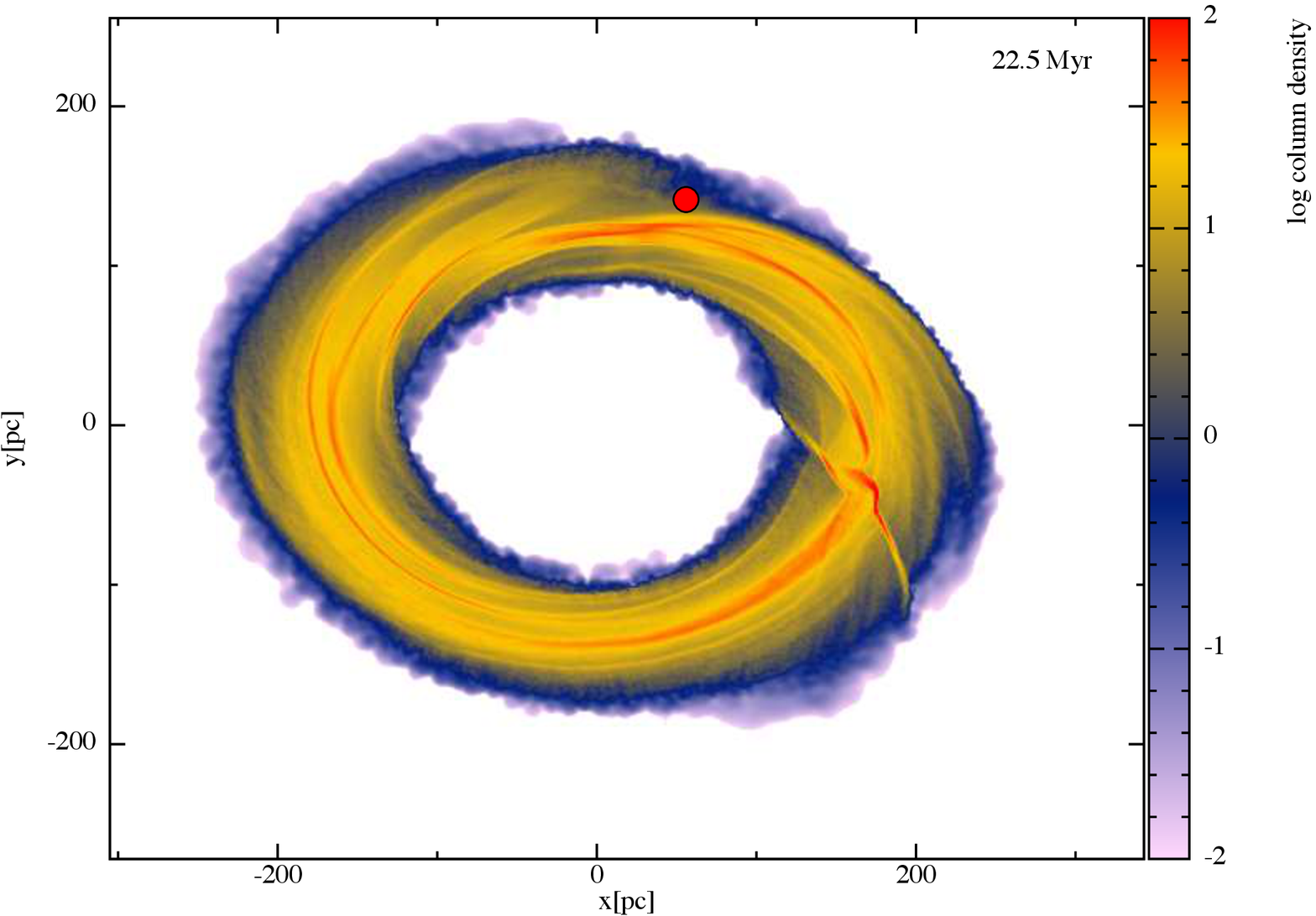}  
\includegraphics[width=3.3in]{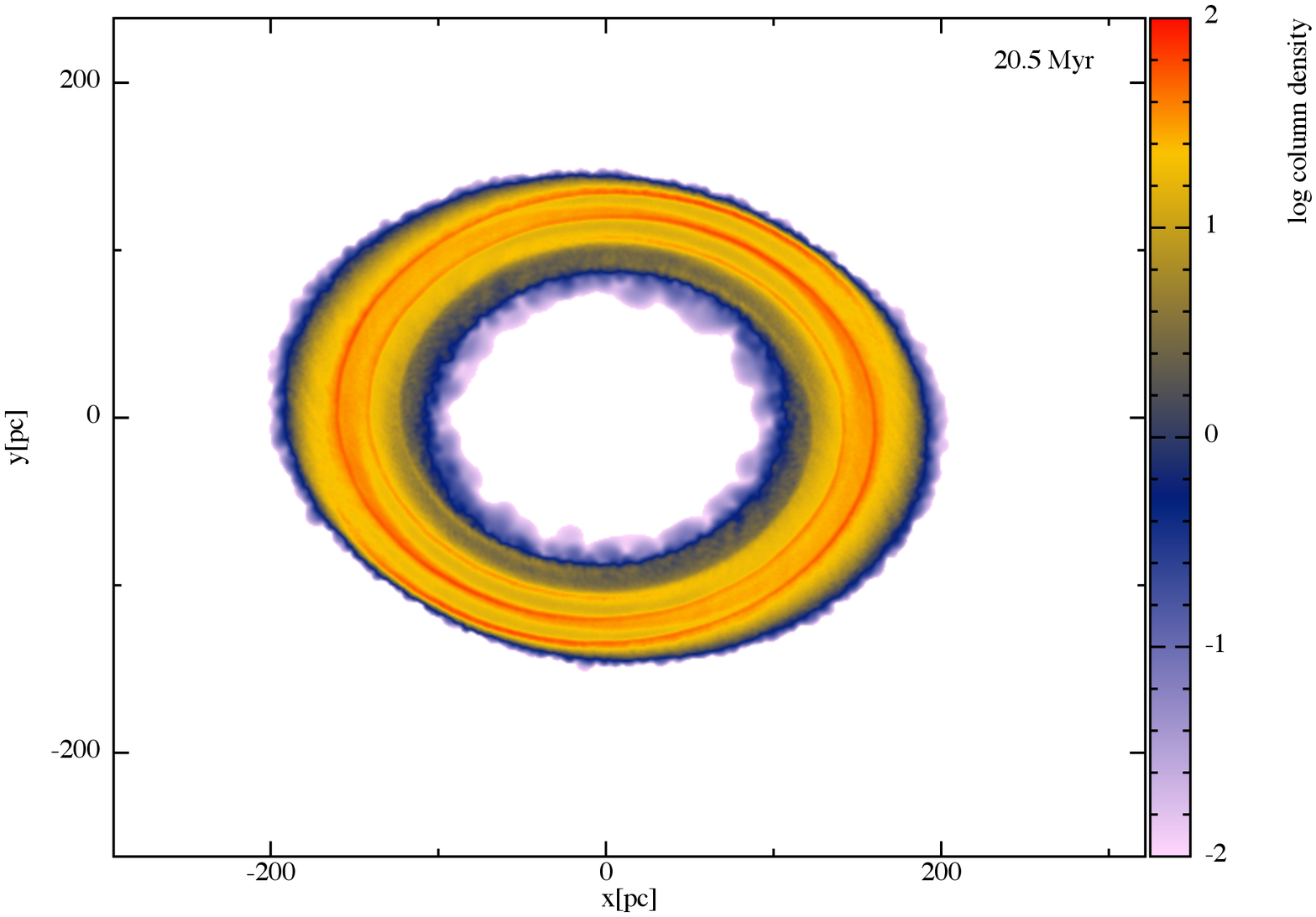}  
\includegraphics[width=3.3in]{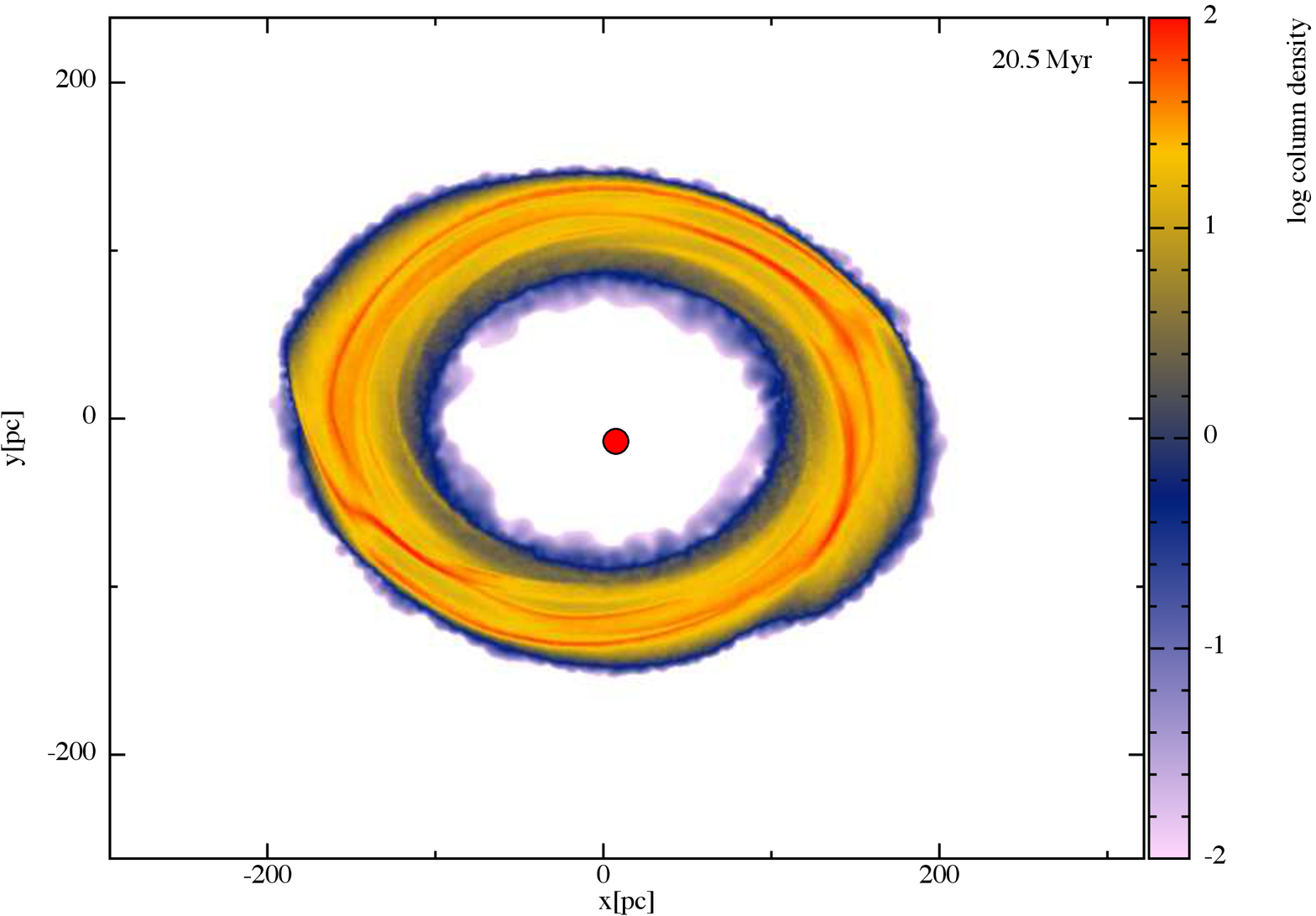} 
 \caption{Face-on column density map of the gas in the control simulations without satellite (left panels) and fiducial simulations with satellite, represented by a red circle (right panels). The top two figures correspond to 2-d models and the bottom two to the 3-d models. The 3-d models show more signs of being perturbed, but the overall CMZ structure remains roughly the same. The CMZ gas is perturbed by the satellite, but mostly remains within its original radial range, producing little inflow.}

   \label{fig1}
\end{center}
\end{figure*}

\begin{figure*}
\begin{center}
\includegraphics[width=3.4in]{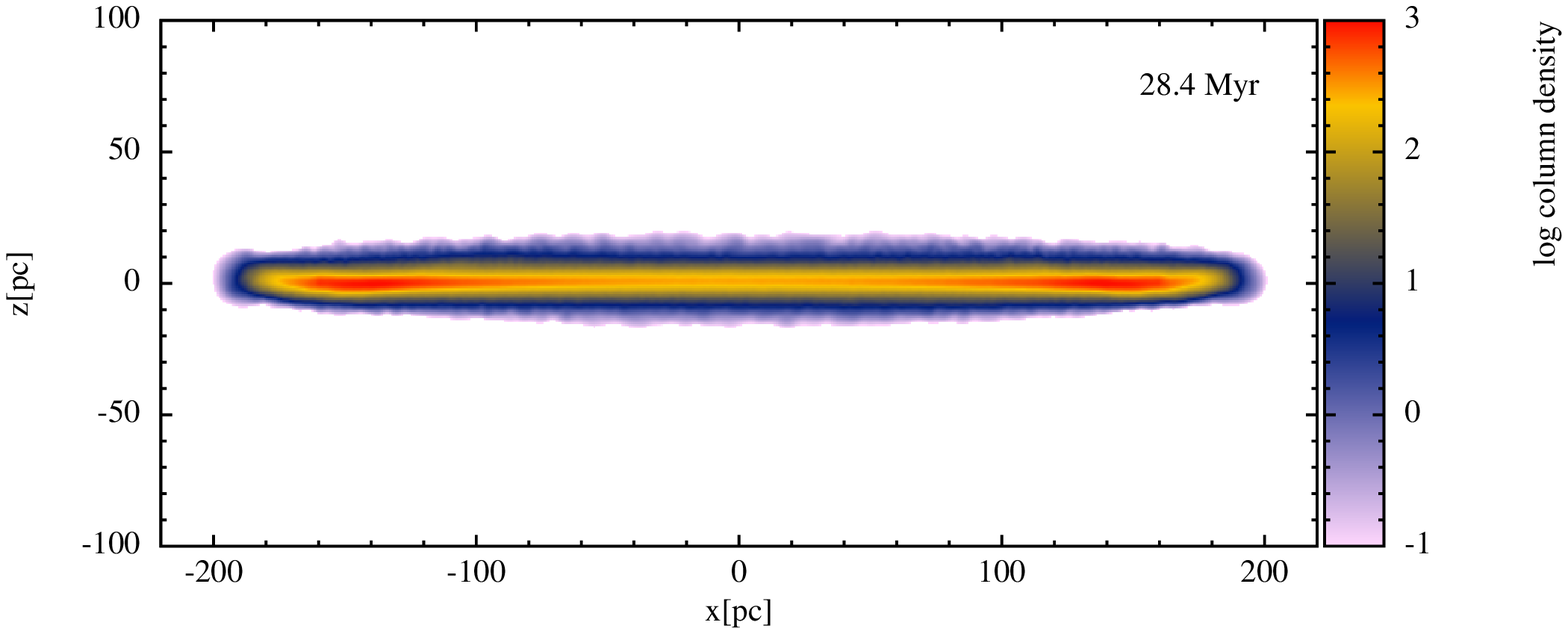}  
\includegraphics[width=3.4in]{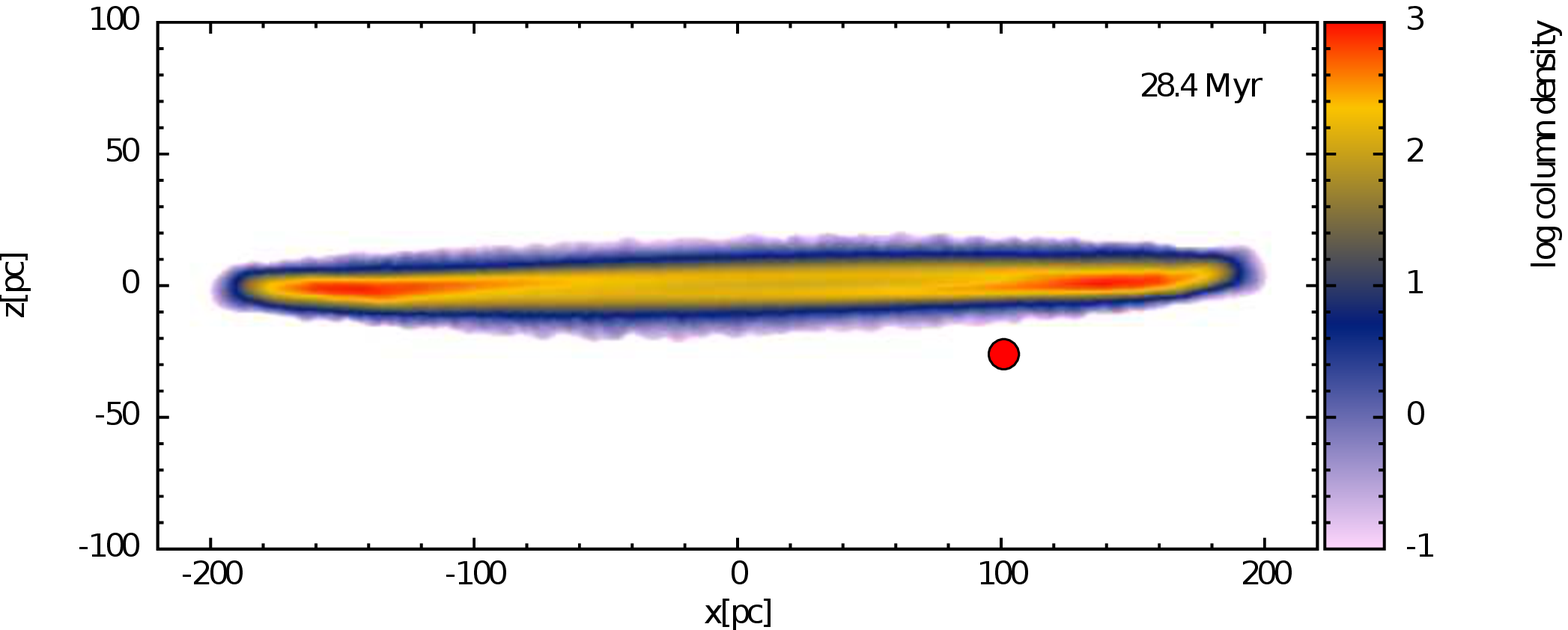} 
 \caption{Edge-on column density map of the gas in the 3-d simulations without satellite (left) and with a satellite in an initial orbit outside the CMZ plane (right), with initial inclination $i=10^\circ$. The visible effect of the satellite is a small warp in the gas distribution, mainly in the innermost orbits.}
   \label{fig2}
\end{center}
\end{figure*}

\begin{table*}
\centering
\caption{Summary of all simulations presented in this work. Column 1 lists the defining characteristic of each model, i.e., the parameter which was modified compared to the fiducial simulation. Column 2 states if the simulation is 2- or 3-dimensional. Columns 3 and 4 show the mass of gas present within 80 pc of the GC, after 20 Myr and 40 Myr respectively.}
\label{my-label}
\begin{tabular}{|l|c|r|r|}
\hline
\hline
      Model characteristics & 2d/3d & M$_{20 \,\rm{Myr}}$ [$\rm M_\odot$] & M$_{40 \,\rm{Myr}}$ [$\rm M_\odot$] \\ 
\hline    
\hline    
    Control model without satellite    & 2d &  0 & 0 \\ 
    & 3d & 341 & 1143  \\ \hline 
   Control model, with $3\times$ gas particles ($N=1.1\times10^6$)  & 2d &  0 & 0 \\ \hline

  Fiducial model & 2d  & 0 & 92 \\ 
   & 3d &  662 & 2688 \\ \hline
 
    Higher satellite mass $2\times10^7\,\rm M_\odot$ & 2d  & 285 & 5161 \\ \hline
     
      Lower satellite mass $2\times10^5\,\rm M_\odot$ & 2d  & 0 & 0 \\ \hline
     
      Counterrotating orbit & 2d & 0 & 57 \\
        & 3d & 481 & 1123 \\ \hline

      Non-coplanar orbit $i=10^\circ$& 3d  & 401 &  1564 \\ \hline
      
      Non-coplanar orbit $i=20^\circ$& 3d  & 441 &  1384 \\ \hline
      
      Non-coplanar orbit $i=40^\circ$& 3d  & 521 &  1624 \\ \hline
      
      Satellite initial velocity $200\,\rm{km/h}$& 3d  & 542 &  2086 \\ \hline
      
      Satellite initial velocity $50\,\rm{km/h}$& 3d  & 662 &  2808 \\ \hline
      
      Lower gas temperature $T=100\,\rm K=0.1\,T_0$ & 3d  & 762 & 3149 \\ \hline
     
      Higher gas temperature $T=8970\,\rm K$, $Q=3\,Q_0$ & 3d  & 481 & 2889 \\ \hline
      
      Lower bar pattern speed of $30\, \mathrm{km}\, \mathrm{s}^{-1}\, \mathrm{kpc}^{-1}$  & 3d &  426 & 1534 \\ \hline
\end{tabular}
\end{table*}

\subsection[]{Longer-term evolution}

During the last 10 Myr in both our 3d fiducial simulation and the control simulation without satellite, we observe an abrupt change in the orbits of part of the inner ring.  There is a significant vertical velocity component, which leads to a stream-like infall of gas of the order of $10^6\,\rm M_\odot$. 
We decided to explore the further evolution of the CMZ ring by running our 3d simulations for an additional 100 Myr. We find the same instability evolution regardless of the presence of the satellite, therefore, we attribute it to the properties of the Galactic potential: being tri-axial, the non negligible vertical component of the potential produces a nutation starting from the innermost orbits and propagating through the CMZ ring \citep{Jeon09}. Even though this is an interesting effect that could potentially produce mass infall and star formation, to our knowledge there have been no detection of stream-like molecular inflows in the Galactic Center.
Further work is needed to explore this longer-term effect further, which we defer to a future paper.

\section{Conclusion}

We presented numerical models of the interaction of the core of a satellite galaxy with the gas in the CMZ.  In all our simulations only a very small fraction of the gas ($\sim10^3\,\rm M_\odot$) reaches the inner $80\,\rm{pc}$ in the 40 Myr after the first satellite collision, and a completely negligible amount reaches the more interesting inner scales (10--30 pc). 
This result holds within our exploration of the parameter-space for the system, such as the CMZ ring stability, the presence of the satellite galaxy, and its properties.
In conclusion, the satellite perturbation does \emph{not} seem sufficient to produce neither star clusters at 30 pc, nor nuclear activity as proposed by L13. In the general context of hierarchical galaxy evolution, our results hint to a low efficiency of gas inflow from interactions between satellites and central molecular zone analogues.

\section*{Acknowledgments}
 
We thank Sungsoo Kim for very helpful suggestions on the  setup of the initial conditions and the Galactic potential, Pau Amaro-Seoane for useful discussions on the shape of the 3D potential,  Victor Debattista for the discussions on the recent studies of the Milky Way bar pattern speed, Germ\'an G\'omez-Vargas for discussions on the Fermi Bubbles, and Jihye Shin for sharing her new draft on CMZ simulations.
We also acknowledge Tamara Bogdanovi\'c, M\'arcio Catelan, Kelly Holley-Bockelmann, and Nelson Padilla for their advice and discussions at the beginning of this project. We thank the anonymous referee for her/his comments and suggestions which contribute to a significant improvement of this study. 

The simulation figures were produced with the public code {\sc splash} \citep{Price07}.

This research was partially funded by CONICYT-Chile through FONDECYT (1141175), Basal (PFB0609) and Anillo (ACT1101) grants.
 The simulations presented in this letter were performed with the Geryon computer at the Center for Astro-Engineering UC, part of the BASAL PFB-06, which received additional funding from QUIMAL 130008 and Fondequip AIC-57 for upgrades.

\label{lastpage}

\end{document}